\documentclass[reqno,12pt]{iopart}
\usepackage{iopams}

\newcommand{\N}{\mathcal{N}}
\date{}
\begin{document}
\title[ Representations of Coherent and Squeezed States in a $f$-deformed Fock Space]
{Representations of Coherent and Squeezed States in a $f$-deformed Fock Space}
\author{R. Roknizadeh, and M. K. Tavassoly\\{Department of Physics, University of Isfahan, Isfahan,
         Iran}}
\eads{\mailto{rokni@sci.ui.ac.ir},
\mailto{mk.tavassoly@sci.ui.ac.ir}}
\begin{abstract}
   We establish some of the properties of the states interpolating
   between number and coherent states denoted by $| n \rangle_{\lambda}$;
   among them are the reproducing of these states
   by the action of an operator-valued
   function on $| n \rangle$ (the standard Fock space) and the fact
   that they can be regarded as $f$-deformed coherent
   bound states. In this paper we use them, as the basis
   of our new Fock space which in this case are not
   orthogonal but normalized. Then by some special superposition of
   them we obtain new representations for coherent and squeezed
   states in the new  basis. Finally the statistical
   properties of these states are studied in detail.
\end{abstract}

 {\bf keywords}: coherent states, squeezed states,
non-orthogonal basis\\
 \maketitle
\section{{\bf Introduction}}\label{sec-intro}
    As is well-known the Hilbert space is a natural framework for
    the mathematical description of many areas of physics:
    certainly for quantum mechanics (and quantum field theory),
    signal and image analysis [Lynn(1986)], etc. In all of these cases
    any arbitrary vector can be represented in terms of the
    elements of some basis $\{|n\rangle, n \in  \mathbf{N}\}_{n=0}^\infty$. The
    orthonormal basis is the one that mostly advocated by
    mathematicians. But unfortunately, orthonormal basis are
    sometimes difficult to find and hard to work with. So the
    generalization to the non-orthogonal basis vectors (such as
    the one we will introduce in this paper denoted by
    $\{|n\rangle_\lambda, n \in \mathbf{N}\}_{n=0}^\infty$) has been proposed.
    These vectors have the properties such as fast convergence,
    uniqueness of decomposition, etc. The resulting object is
    called a {\it "frame"}[Ali(1993)]; or more precisely a {\it "discrete frame"}.
    The non-orthogonal states are also used as a tool in {\it"generalized measurement"}
    that is a very important subject in view of foundations of
    quantum mechanics, quantum nondemolition measurement and quantum
    information theory[Peres(1995)].
    The present work which contains the construction of coherent
    states(CSs) and squeezed state(SSs) using a special set of non-orthogonal but normalizable basis
    $\{|n\rangle_\lambda, n \in \mathbf{N}\}_{n=0}^\infty$ instead of
    orthonormal one $\{|n\rangle, n \in \mathbf{N}\}_{n=0}^\infty$ can be
    classified in the discrete frame background. Thus in our
    opinion this approach provides a more fundamental and certainly
    more flexible rule than that of the usual one considered extensively in
    the physical literature for construction of new generalized
    CSs, the concept which have attained the most applications
    in quantum optics, as well as other fields of physics[Klauder(1985)-Perelomov(1986)].

    For simple harmonic oscillator (SHO) in natural unit and with
    unit frequency we have
\begin{equation}\label{HamiltSHO}
    H= a^{\dag} a + \frac{1}{2}, \qquad  [a, a^{\dag}]=1, \qquad
   (a^\dag)^{\dag}=a,
\end{equation}
    where $a$, ${a^\dag}$ and $H$ are standard annihilation, creation
    and Hamiltonian operators, respectively. Following the particular kind of deformation proposed by
    J. Beckers {\it et al} [Beckers(1998)] we consider the parametric harmonic
    oscillator by deforming the creation operator in such a way that
\begin{equation}\label{def-creation}
    a^{\dag}_{\lambda} = a^{\dag} + {\lambda} I, \qquad   \lambda \in
    \mathbf{R},
\end{equation}
    without changing the annihilation operator. So the
    $\lambda$-Hamiltonian becomes
\begin{equation}\label{def-Hamilt}
   H_\lambda=a^\dag_{\lambda} a + \frac{1}{2}.
\end{equation}
   Note that $(a^\dag_\lambda)^\dag \neq a_\lambda, H_\lambda^\dag
   \neq H_\lambda$, but as for SHO yet we have:
\begin{equation}\label{comut}
   [a, a^\dag_\lambda]=1, \qquad
   [H_\lambda, a^\dag_ \lambda]=a^\dag_ \lambda.
\end{equation}
    To  manifest our motivation for considering the special kind
    of deformation in Eq. (\ref {def-creation}) we offer the following discussion.
    In a previous paper we have shown that a large class of
    generalized CSs can be obtained by changing the basis in the
    underlying Hilbert space. We have presented a systematic
    formalism that the particular deformation which we employed in
    the present paper is a special case of the general  scheme
    have been introduced in [Ali(2004)]. To clarify more, we will explain
    briefly the setting. Let $ \mathcal{H} $ be a Hilbert space and
    $T, T^{-1}$ be operators densely defined and closed on
    $ \mathcal{D}(T)$ and $\mathcal{D}(T^{-1})$,
    respectively and $F=T^\dag T$. Two new Hilbert spaces $\mathcal{H}_F, \mathcal{H}_{F^{-1}}$
    are the completions of the sets $ \mathcal{D}(T)$ and $\mathcal{D}(T^ {\dag^{-1}})$
    with the scalar products
\begin{equation}\label{scalar}
    \langle f|g \rangle_F = \langle f|Fg \rangle_{\mathcal{H}}, \qquad
    \langle f|g \rangle_{F^{-1}} = \langle f|F^{-1}g
    \rangle_{\mathcal{H}},
\end{equation}
    respectively. Considering the generators of the
    Weyl-Heisenberg algebra as  bases on $\mathcal{H}$, we may obtain the
    transformed generators on $\mathcal{H}_F$ such as:
\begin{equation}\label{genHF}
    a_F=T^{-1}a T, \qquad a^{\dag}_F = T^{-1}a^\dag T, \qquad
    N_F=T^{-1} N T.
\end{equation}
    A similar argument may be followed for the Hilbert space
    $ \mathcal{H}_{F^{-1}}$, which properties were discussed in
    [Ali(2004)]. It is obvious that the oscillator algebra remains unchanged.
    Now choosing $T=\exp({-\lambda a})$ and using (\ref {genHF}) for the annihilation and
    creation operators on $\mathcal{H}_F$ we recover the exact form of
    the deformation introduced by H. Fu {\it et al}[Fu(2000)] and J. Beckers {\it et al} [Beckers(1998)]
    which were employed in this paper, {\it i.e.}
\begin{equation}\label{def-algebra}
    a_F=a, \qquad a_F^{\dag}=a^\dag +\lambda I.
\end{equation}
    Other selections for $T$-operator lead to other families of generalized
    CSs.  In this manner we have established the basic place of
    the particular kind of deformation, we have used in this work,  in the general theory of
    constructing the CSs in non-orthogonal basis. The case in which the annihilation
    operator is deformed as $a_\lambda = a+\lambda I$,
    also have been already considered and discussed by S. T. Ali and us in [Ali(2004)].

    The eigenvalue equation
\begin{equation}\label{eigen-s}
    H_\lambda|n \rangle_\lambda=E_{n,\lambda}|n \rangle_\lambda,
\end{equation}
    has been solved in [Beckers(1998)], and led to the
    $\lambda$-states
\begin{equation}\label{wave-f}
   \psi_{n,\lambda}(x)=2^n n! \pi^{-1/4}e^{-x^2/2} L_n^{(0)} (-\lambda^2) H_n(x+\lambda /
   \sqrt{2}),
\end{equation}
   where $(L_n^{(0)}$ and $H_n$ are respectively the well-known Laguerre and Hermite polynomials
   of order $n$) and $ H_ {\lambda} $ is isospectral with
   $H_{_{SHO}}$, {\it i.e.}
\begin{equation}\label{eigen-v}
   E_{n, \lambda } = E_n = n+\frac{1}{2},\qquad   n=0, 1, 2, \dots .
\end{equation}
   Also the operation of $a$ and $ a^ {\dag} _ {\lambda} $ on
   $\lambda$-states are as follows:
\begin{equation}\label{anihilate}
   a|n\rangle_\lambda=\sqrt{n}
 \left(\frac{L_{n-1}^{(0)}(-\lambda^2)}{L_n^{(0)}(-\lambda^2)}\right)^{1/2}|n-1\rangle_\lambda
\end{equation}
\begin{equation}\label{create}
  a_{\lambda}^\dag|n\rangle_\lambda=\sqrt{n+1}\left(\frac{L_{n+1}^{(0)}(-\lambda^2)}
  {L_n^{(0)}(-\lambda^2)}\right)^{1/2}|n+1\rangle_\lambda.
\end{equation}
  The states introduced in Eq. (\ref {wave-f}) are not
  orthogonal and their inner products read as:
\begin{equation}\label{inner-p}
  _\lambda \langle m | n
   \rangle_\lambda=[L_m^{(0)} (-\lambda^2)
   L_n^{(0)}(-\lambda^2)]^{-1/2}\sum_k \frac{\lambda^{2k+m-n}(n!m!)^{1/2}}{k!(n-k)!(m-n+k)!}
\end{equation}
   but it is trivial from Eq. (\ref{inner-p}) that they are
   normalized, {\it i.e.} $_\lambda \langle n | n
   \rangle_\lambda=1$
\section{{\bf Some remarkable points}}\label{sec-intro}
   Before we offer our main work, let us list some of the interesting and important points we may
   conclude. The following results may be considered as a
   complementary in regard to the previous related works mentioned
   earlier[Fu(2000) and Beckers(1998)].

   {\bf a}) It can easily be shown that the states
   $|n\rangle_\lambda$ can be regarded as the basis for our new
   Fock space, since the necessary and sufficient conditions
   mentioned for a one-dimensional quantum mechanical Fock space in
  [Bardek(2000)] will be satisfied. In our problem these conditions are briefly
  {\bf i)} existence of a vacuum state such that $a|0\rangle =0$, {\bf ii)}
  $\langle 0|aa^\dag _\lambda |0\rangle > 0$, {\bf iii)} $[aa^\dag_\lambda,
  a^\dag_\lambda a]\neq 0$ and $aa^\dag_\lambda \neq a^\dag
  _\lambda a $.

  {\bf b)} Despite of the appearance of Eq. (\ref {inner-p}),
   $_\lambda\langle m|n \rangle_\lambda =_\lambda\langle n|m \rangle_\lambda$ holds
   in the Hilbert space whose basis spanned by states $ |n\rangle_
   \lambda $.

  {\bf c)} $a^{\dag}_\lambda a|n\rangle_\lambda \equiv \hat{n}_\lambda
   |n\rangle_\lambda=n|n\rangle_\lambda$; so $\hat{n}_\lambda$ can
   be thought as number operator in the new Fock space. Moreover
   from this equation we see that $(\hat{n}+\lambda
   a)|n\rangle_\lambda = n |n\rangle_\lambda$ which indicates simply the ladder operator
   formalism [Wang(2000)] of the state $|n\rangle_\lambda$.

  {\bf d)} By Eqs. (\ref{anihilate}) and (\ref{create}) we obtain respectively
  \begin{equation}\label{iter-ani}
    a^n |n\rangle_\lambda=\left(\frac{n!}
   {L_n^{(0)}(-\lambda^2)}\right)^{1/2}|0\rangle_\lambda,
\end{equation}
    and
\begin{equation}\label{iter-crea}
    |n\rangle_\lambda=\frac{(a^\dag_\lambda)^n}{\left(n!L_n^{(0)}(-\lambda^2)\right)^{1/2}}|0\rangle_\lambda.
\end{equation}
   Since $|0\rangle_\lambda=|0\rangle_{_{SHO}}$, using Eqs. (\ref {def-creation}) and
   (\ref {def-Hamilt}) and Binomial formula we get easily the relation between
   the standard and $\lambda$-Fock space as:
\begin{equation}\label{exp-nlamb-n}
  |n\rangle_\lambda = \sum_{m=0}^{n} \frac{(n!)^{1/2}\lambda^{n-m}}{(n-m)!
  \left( m!L_n^{(0)}(-\lambda^2)\right)^{1/2}} |m\rangle,
\end{equation}
  which suggests that every state $|n\rangle_\lambda$ in this
  non-orthogonal Hilbert space can be regarded as a special finite
  superposition of $|0\rangle,|1\rangle,...,|n\rangle$ in the standard
  Fock space $|n\rangle_{_{SHO}}$ (eliminated indices SHO).

 {\bf e}) As another point, we see that according to the result of
   our calculation in Eq. (\ref{exp-nlamb-n}), the states $|n\rangle_\lambda$ are
   indeed the intermediate number-coherent states denoted by $\|\eta,
   n\rangle$ which have been already derived by H. Fu {\it et al} in [Fu(2000)]:
\begin{equation}\label{dis-exit}
    |n\rangle_\lambda \equiv \|\eta, n\rangle=D(-\lambda)|\lambda,
    n\rangle,
\end{equation}
   where $\lambda=\sqrt{\frac{1-\eta}{\eta}}$ and $\eta$ is a real probability  restricted to
   $0 < \eta \leq1$, $D(-\lambda)$ is the ordinary displacement operator will
   be defined later in Eq. (\ref {displacement}) and
\begin{equation}\label{lambda-n}
       | \lambda, n\rangle =
        \frac{1}{\sqrt{n!L_n^{(0)}(-\lambda^2)}} a^{\dag^n}|\lambda \rangle
\end{equation}
   is the photon-added coherent state or exited coherent states and
   $|\lambda\rangle=D(\lambda)|0\rangle$ is a coherent
   state. With this respect H. Fu {\it et al} suggested that the states
   $\|\eta,n\rangle$ (and also $|n\rangle_\lambda)$ are displaced exited
   coherent states. We are thankful for suggestion a scheme in Ref. [Fu(2000)]
   that during an experiment one can generate these states. They
   have shown that $\lambda$-parameter depends physically on an
   external driving field ($A$) in a cavity and the cavity resonant frequency ($\omega$)
   through the relation $\lambda=A/\omega$.

{\bf f}) Using Eqs. (\ref {dis-exit}) and (\ref{lambda-n}) and
   BCH lemma, the rhs of Eq.
   (\ref {exp-nlamb-n}) can be replaced by the compact and interesting
   formula:
\begin{equation}\label{T-op}
   |n\rangle_\lambda=\frac {e^{\lambda a}}{\sqrt{L_n^{(0)}(-\lambda^2)}} |n\rangle \equiv
   T_{n, \lambda}|n\rangle,
\end{equation}
   where $T_{n, \lambda}$ is an operator-valued function which
   exponentially depends on $\lambda$ and the annihilation
   operator. The coefficient $[L_n^{(0)}(-\lambda^2)]^{-1/2}$ is a
   $c$-number which keeps the normalizability of the deformed Fock
   space, and the order of the Laguerre polynomial match with the
   old number state which we wish to transform.

   {\bf g}) Recently  Shanta {\it et al} [Shanta(1994)] and in
   a systematic manner Man'ko {\it et al} [Manko(1996)]
   introduced nonlinear coherent
   states as eigenfunctions of a deformed annihilation
   operator $A=a f(n)$, such that
\begin{equation}\label{ani-def}
   A|\alpha, f\rangle = \alpha |\alpha, f\rangle,
\end{equation}
   which the representation of these states in $|n\rangle$ basis
   is giving by
\begin{equation}\label{nonl-cs}
   |\alpha, f\rangle = \N_f (|\alpha|^2)\sum_{n=0}^{\infty} \frac {\alpha^n}{\sqrt n f(n)!}|n\rangle
\end{equation}
   where $f(n)$ is the nonlinearity function,
   $f(n)! \doteq f(0)f(1)...f(n)$, (by convention $f(0) = 1$) and $\N_f$
   is an appropriate normalization factor. More recently
   J. Re'camier {\it et al} in [Re'camier(2003)] proposed a definition for
   $f$-coherent bound states as
\begin{equation}\label{nonl-Bcs1}
   |\alpha, f\rangle_B = \N_{f, \alpha}^{(m)} \sum_{n=0}^m \frac
   {\alpha^n}{\sqrt n f(n)!}|n\rangle,
\end{equation}
   where $\N_{f,\alpha}^{(m)}$ is a normalization factor. Following
   these consideration it is possible to rewrite the states $|n\rangle_\lambda$
   in the form of normalized $f$-coherent bound states
\begin{equation}\label{nonl-Bcs2}
  |m\rangle_\lambda \equiv \parallel \eta, m \rangle =
  \left(\frac{\lambda^{2m}m!}{L_m^{(0)}(-\lambda^2)}\right)^{1/2}
  \sum_{n=0}^m \frac{(\lambda^{-1})^n}{\sqrt{n!}(m-n)!}|n\rangle
  = |\lambda^{-1}, f\rangle_B.
  \end{equation}
   So the space spanned by $|n\rangle_\lambda $ can be regarded as
   a nonlinear (or $f$-deformed) Fock space, with the nonlinearity
   function $f(\hat{n})=(m-\hat{n})$. It must be noted that the
   states $|n\rangle_\lambda$ are not eigenstates of the
   annihilation operator $a$, because of the finiteness of the
   upper bound of the summation.
\section{{\bf Construction of coherent states in $|n\rangle_\lambda$ basis}}\label{sec-relat}
    For constructing the coherent states we begin with the definition
    of CSs as eigenstates of annihilation operator [Manko(1996), Das(2002)]
\begin{equation}\label{ani-def-nl}
   a|\alpha, \lambda \rangle = \alpha|\alpha,\lambda\rangle; \qquad
   \alpha\in \mathcal{C},
\end{equation}
   where we call $|\alpha, \lambda\rangle$, $\lambda$-CSs. Now if we
   expand $|\alpha, \lambda\rangle$ in terms of
   $|n\rangle_\lambda$ basis
\begin{equation}\label{lamb-cs}
  |\alpha, \lambda \rangle = \sum_{n=0} ^\infty C_n | n \rangle_{\lambda},
\end{equation}
   then by Eqs. (\ref{anihilate}) and (\ref{ani-def-nl}) we get
\begin{equation}\label{20}
  \sum_{n=0}^\infty C_n\sqrt{n}
  \left(\frac{L_{n-1}^{(0)}(-\lambda^2)}{L_n^{(0)}(-\lambda^2)}\right)^{1/2}
  |n-1\rangle\lambda = \alpha \sum_{n=0}^\infty
  C_n|n\rangle_\lambda,
\end{equation}
   from which we find that the coefficients $C_n$ satisfies the
   following recurrence relation
\begin{equation}\label{21}
   C_n = \frac{\alpha^n \left(L_n^{(0)}(-\lambda^2)\right)^{1/2}}{\sqrt{n!}} C_0.
\end{equation}
   The normalization condition of the state $|\alpha, \lambda\rangle, {\it
   i.e.} \langle\alpha, \lambda|\alpha, \lambda\rangle = 1 $ leads to
   a complicated series for coefficient $C_0$ which after some
   straightforward but lengthy calculation one obtains:
\begin{equation}\label{22}
   C_0 = \exp \left[-\lambda \Re
   (\alpha)-\frac{|\alpha|^2}{2}\right],
\end{equation}
   where $\Re(\alpha)$ is the real part of $\alpha$. Finally
   the normalized $ \lambda $-CS takes the form:
\begin{equation}\label{lambda-CS}
   |\alpha, \lambda\rangle =
   \exp \left[-\lambda \Re (\alpha)-\frac{|\alpha|^2}{2}\right]\sum_{n=0}^\infty
   \frac {\alpha^n\left(L_n^{(0)} (-\lambda^2)\right)^{1/2}} {\sqrt{n!}} |n\rangle_\lambda,
\end{equation}
    which is a {\it "new representation"} of canonical CS in
    $|n\rangle_\lambda$ basis. Their inner product which allows
    overcompleteness is
\begin{equation}\label{over-complet}
  \langle \alpha, \lambda | \beta, \lambda \rangle =
  \N_{\alpha, \beta, \lambda} \sum_{m=0}^{\infty} \sum_{n=0}^{\infty}
  \alpha^{*m} \beta^n \sum_k \frac {\lambda^{2k+m-n}}{k! (n-k)! (m-n+k)!}
\end{equation}
   with
\begin{equation}\label{coef}
   \N_{\alpha, \beta, \lambda}=\exp \left[-\lambda \Re(\alpha)-\lambda
   \Re(\beta)-\frac{|\alpha|^2}{2}-\frac{|\beta|^2}{2}\right].
\end{equation}
    Now we imply that by Eq. (\ref {iter-crea}) the $\lambda$-CSs in Eq. (\ref {lambda-CS}) can
    be expressed in terms of the lowest eigenstate of $H_\lambda$ as:
\begin{equation}\label{repres1}
    |\alpha, \lambda\rangle = \exp \left[-\lambda \Re(\alpha)-
    \frac{|\alpha|^2}{2}\right] e^{\alpha
    a_{\lambda}^\dag}|0\rangle.
\end{equation}
   Since $a|0\rangle=0$, by BCH lemma Eq. (\ref {repres1}) can
   be rewritten as
\begin{equation}\label{Lamb-CS-Vaccum}
      |\alpha, \lambda \rangle = e^{-\lambda \Re(\alpha)}
      \exp (\alpha a_{\lambda}^\dag-\alpha^*a )|0\rangle
\end{equation}
   and finally using Eq. (\ref {def-creation}) we have
\begin{equation}\label{displacement}
      |\alpha, \lambda \rangle = e^{i \lambda \Im (\alpha)}
      \exp (\alpha a^{\dag}-\alpha^* a )|0\rangle
      = e^{i \lambda \Im (\alpha)}
      D(\alpha)|0\rangle \equiv D_\lambda (\alpha)|0\rangle
\end{equation}
   where the imaginary part of $\alpha$ is denoted by $\Im (\alpha)$.
   Since $D(\alpha)|0\rangle=|\alpha\rangle$ which is the usual
   CSs, we conclude that $|\alpha,\lambda\rangle$ is identical to
   $|\alpha\rangle$, up to a phase factor equal to  $e^{i \lambda \Im(\alpha)}$, and
   therefore $|\alpha, \lambda\rangle = |\alpha\rangle$ whenever $\alpha \in \mathcal{R}$; a
   result that may be expected from the eigenvalue equation (\ref {ani-def-nl}).
   So obviously there is no problem with resolution of the identity
\begin{equation}\label{resol}
  \int d\mu (\alpha) |\alpha, \lambda \rangle \langle \alpha,
  \lambda|=I.
\end{equation}
   We would like to emphasize that all we have down in this
   section is obtaining the explicit form of canonical CS in a
   deformed Fock space $|n \rangle_\lambda$, which in this case are
   nonorthogonal, and we called them new representation
   of canonical CS (Eq.(\ref {lambda-CS})). In fact $|\alpha, \lambda\rangle$ and
   $|\alpha\rangle$ belong to the same ray in the projective Hilbert
   space (we will explain more about this result in the appendix). As
   another result we conclude that by some particular superpositions of displaced exited
   coherent state $|n\rangle_\lambda$ (which exhibit squeezing [Fu(2000)]), we have obtained
   canonical CSs.
\section{{\bf Time evolution of $\lambda$-coherent states}}\label{sec-displce}
     We are able to consider the dynamical evolution of the
     $\lambda$-CSs, in $|n\rangle_\lambda$ basis, which is simply obtained,
     because the spectrum of $H_\lambda$ is the same as
     $H_{_{SHO}}$:
\begin{eqnarray}\label{temporal}\nonumber
  U(t)|\alpha, \lambda \rangle
  &=& \exp \left[-\lambda
  \Re(\alpha)-\frac{|\alpha^2|}{2}\right]
  \sum_{n=0}^{\infty}\frac{\alpha^n}{\sqrt{n!}}\left( L_n^{(0)}(-\lambda
  ^2)\right)^{1/2}e^{-iH_\lambda t}|n\rangle_\lambda  \nonumber\\
  &=& \exp\left[-\lambda
  \Re(\alpha)-\frac{|\alpha^2|}{2}\right]
  \sum_{n=0}^{\infty}\frac{\alpha^n}{\sqrt{n!}}\left( L_n^{(0)}(-\lambda
  ^2)\right)^{1/2}e^{-i(n+\frac{1}{2})t}|n\rangle_\lambda \nonumber\\
  &=&e^{-it/2}|\alpha(t), \lambda \rangle,
\end{eqnarray}
     where we have used $\alpha(t)\equiv \alpha e^{-it}$. This means that the time
     evolution of $\lambda$-CSs in this nonorthogonal basis remain
     coherent for all times ({\it temporal stability}).
\section{{\bf Statistical properties of $|\alpha, \lambda\rangle$ in $|n\rangle_\lambda$ bases}}\label{sec-displce}
   For standard CS $|\alpha\rangle$, the occupation
   number distribution is Poissonian
\begin{equation}\label{Poisson}
   P(n)=\left |\langle n|\alpha\rangle
   \right|^2=e^{-|\alpha|^2}\frac{|\alpha|^{2n}}{n!},
\end{equation}
   whose mean and variance are equal to $|\alpha|^2$. Similarly in
   the case of our $\lambda$-CSs, we define:
\begin{equation}\label{32}
   P_\lambda(m)=| _\lambda\langle m|\alpha, \lambda \rangle|^2 \\
               =\frac {m!}{L_m^{(0)}(-\lambda ^2)}e^{-|\alpha|^2}
               \sum_{n=0}^m \sum_{p=0}^m \frac{\lambda^{2m-n-p} \alpha^{*p}\alpha^n}
               {p!(m-p)!n!(m-n)!}
\end{equation}
     as the probability of finding the states $|\alpha,\lambda
     \rangle$ in $|n \rangle_\lambda$ basis. It is easy to verify
     that $\lim_{ \lambda \rightarrow 0} P_\lambda(m)=P(m)$. While the
     distribution in $ |n\rangle_{_{SHO}} $ basis is Poissonian, this
     is not so in $ |n\rangle_\lambda $ basis. As usual we have:
\begin{equation}\label{rr}
   _ \lambda\langle \hat{m} \rangle_\lambda = \sum_m
   m P_\lambda(m),\qquad
   _\lambda\langle {\hat{m}}^2 \rangle_\lambda = \sum_m m^2P_\lambda(m).
\end{equation}
   To check Poissonian, sub-Poissonian or supper-Poissonian statistics, we can
   evaluate Mandel parameter defined as:
\begin{equation}\label{mandel}
  Q(\alpha,\lambda) =
  \frac  {\langle{\hat{m}}^2\rangle-{\langle{\hat{m}\rangle}}^2}
  {\langle\hat{m}\rangle}-1.
  \end{equation}
   We see from figure (1) that the states $ |\alpha, \lambda \rangle
   $ exhibit sub-Poissonian (nonclassical) or supper-Poissonian
   statistic in $|n\rangle_\lambda$ basis depending on the values of
   $\alpha$. While it is seen that for $\alpha = 1, 2$ the state is
   sub-Poissonian, when $\alpha = -1, -2$ it is supper-Poissonian.
   As another feature, our numerical results show that when $\alpha \in \mathcal{R}$ and $\alpha < 0
   $ for large values of $\lambda $, the statistical
   behavior of the state is nearly close to Poissonian.
\section{{\bf Squeezed states in terms of $|n\rangle_\lambda$-bases}}\label{sec-ss}
   According to the statement of A. I. Solomon and J. Katriel
   [Solomon(1990)], the conventional squeezed states are obtained by the
   action of a linear combination of creation and annihilation
   operators on an arbitrary state.
   Now by generalizing this procedure to $a^\dag _\lambda = a^\dag +\lambda
   I$ and $a_\lambda=a$ of the deformed oscillator algebra with $a, a^\dag_\lambda$ as
   annihilation and creation operators we have:
\begin{equation}\label{SSs}
  (a-\xi a^\dag_\lambda)|\xi, \lambda \rangle =0, \qquad \xi \in
  \mathcal{C}.
\end{equation}
   This equation for $\lambda=0$ leads to the squeezed vacuum
   states
\begin{equation}\label{SS}
   |\xi \rangle = C_0
   \sum_{n=0}^{\infty}\xi^n
   \left[\frac{(2n-1)!!}{(2n)!!}\right]^{\frac{1}{2}}
   |2n\rangle_\lambda =C_0S(\xi)|0\rangle,
\end{equation}
   where the normalization coefficient may be determined as:
\begin{equation}\label{}
  C_0=\left[\sum_{n=0}^{\infty}\frac{|\xi|^{2n}(2n-1)!!}{(2n)!!}\right]^{-\frac{1}{2}},
\end{equation}
    and $S(\xi)=\exp\left[\frac{\xi(a^\dag)^2}{2}\right]$ is the
    squeezed operator. But in general from Eq. (\ref {SSs}) and by the
    same procedure we have done above and in section {\bf2}, for
    $\lambda\neq 0$ we will arrive at a {\it "new representation"} for
    squeezed state ($\lambda$-SSs) as:
\begin{equation}\label{represSSs}
   |\xi, \lambda \rangle = C_0
   \sum_{n=0}^{\infty}\xi^n
   \left(L_{2n}^{(0)}(-\lambda^2)\right)^{\frac{1}{2}}
   \sqrt{\frac{(2n-1)!!}{(2n)!!}}|2n\rangle _\lambda, \qquad \xi \in
   \mathbf{D},
\end{equation}
   where $\mathbf{D}=\{\xi \in \mathcal{C}|\quad |\xi| < R \}$ is a
   disk centered at the origin in the complex $\xi$ plane with radius $R$,
   which by a suitable transformation can be transformed to a unit
   disk. For the normalization factor one may get:
\begin{equation}\label{coef}
   C_0=\left[ \sum_{n=0}^\infty\sum_{m=0}^\infty\xi ^n \xi^{\ast
   m}(2n-1)!!(2m-1)!!\sum_k \frac{\lambda
   ^{2k+2m-2n}}{k!(2n-k)!(2m-2n+k)!}\right]^{-\frac{1}{2}}.
\end{equation}
   These $\lambda$-SSs are normalizable provided that the
   coefficients $C_0$ is nonzero and finite. It is very hard if not impossible, to
   obtain the analytic form of the radius of convergence for
   $C_0$. This is due to the presence of
   three sigmas and two distinct parameters $\lambda$ and $\xi$ in
   the series (Eq. \ref {coef}).
   But this has been checked on the disk $\mathbf{D}$ for a large value of $\lambda$ and we
   obtained a finite radius of convergence for $C_0$. But our investigations on numerical results shows that as
   $\lambda$ increased, the radius of convergence reduced and
   this will confine the freedom of choosing $\lambda$.

   We see that these states are also even but in
   $|n\rangle_\lambda$ basis, although if we transform them by Eq.
   (\ref {exp-nlamb-n}) to the ordinary basis $|n\rangle$, no definite parity can
   be observed. To change this representation of squeezed states, Eq.
   (\ref {represSSs}), to the old standard Fock states, we obtain
\begin{equation}\label{}
     |\xi, \lambda\rangle = C_0 \exp \left[
     \frac{\xi(a^\dag_\lambda)^2}{2} \right]|0\rangle = C_0 \exp \left[
     \frac{\xi \lambda^2}{2} \right] S(\xi) D(\xi \lambda)
     |0\rangle,
\end{equation}
    where as defined before $S(\xi)$ and $D(\xi \lambda)$ are
    squeezed and displacement operators, respectively.
    The state $|\xi, \lambda \rangle$ in the
    standard Fock space is the one which is named as squeezed
    coherent state. From the above discussion we immediately conclude that translating $|\xi, \lambda
    \rangle$ in $|n\rangle _\lambda$ basis to a state in the
    standard Fock space $|n\rangle$ does not coincide to the one
    we referred in Eq. (\ref {SS}). This is due to the
    fact that in obtaining the SSs, both of the annihilation and deformed creation operators
    are contributed (Eq. \ref{SSs}). To check the nonclassical
    properties of these states first we investigate the quadrature
    squeezing.
     The quadratures $x$ and $p$ are related to
    $a$ and $a^\dag$ according to:
\begin{equation}\label{xp}
    \hat{x}=\frac {a +a^\dag}{\sqrt 2}, \qquad  \hat{p}=\frac {a - a^\dag}{i \sqrt
    2}.
\end{equation}
    As usual the variances of coordinates $x$ and momentum $p$ are
    as follows:
\begin{equation}\label{delx}
  (\triangle x)^2 = \langle x^2 \rangle - \langle x \rangle ^2
                  = \frac {1}{2} \left[1+\langle a^2\rangle
                   + \langle a^{\dag 2} \rangle + 2 \langle a^\dag a \rangle -
                   \langle a \rangle^2 - \langle a^\dag \rangle^2- 2 \langle
                   a\rangle\langle a^\dag \rangle \right],
\end{equation}
\begin{equation}\label{delp}
  (\triangle p)^2 = \langle p^2 \rangle - \langle p \rangle ^2
                  = \frac {1}{2} \left[1-\langle a^2\rangle
                   - \langle a^{\dag 2} \rangle + 2 \langle a^\dag a \rangle
                   +\langle a \rangle^2 + \langle a^\dag \rangle^2- 2 \langle
                   a\rangle\langle a^\dag \rangle \right],
\end{equation}
 where all of the expectation values should be calculated with respect to
 the states $| \xi, \lambda \rangle$. All we need for evaluating
 the rhs of the Eqs. (\ref{delx}) and (\ref{delp}) are the terms such as
\begin{eqnarray}\label{}\nonumber
   _\lambda \langle m|(a^\dag _\lambda)^k | n\rangle _\lambda=
   (n+k)! \left( \frac {m!}{n!
   L_n^{(0)}(-\lambda^2)L_m^{(0)}(-\lambda^2)}\right)^{\frac{1}{2}}
   \\
   \times \sum_l \frac {\lambda^{(2l+m-n-k)}}{l!(n+k-l)!(m-n-k+l)!}
\end{eqnarray}
\begin{equation}\label{}
   _\lambda \langle m|a^k | n\rangle _\lambda=
   \left( \frac {m!n!}
   {L_n^{(0)}(-\lambda^2)L_m^{(0)}(-\lambda^2)}\right)^{\frac{1}{2}}\sum
   _l \frac {\lambda^{(2l+m-n+k)}}{l!(n-k-l)!(m-n-k+l)!}
\end{equation}
and
\begin{eqnarray}\label{}\nonumber
  _\lambda \langle m| (a^\dag_\lambda)^r a^k|n\rangle_\lambda =
        \frac {(n-k+r)!}{(n-k)!} \left(
        \frac{m!}{L_m^{(0)}(-\lambda^2)L_n^{(0)}(-\lambda^2)}\right)^{\frac{1}{2}}
        \\ \times \sum_l \frac{\lambda
        ^{2l+m-n+k-r}}{l!(n-k+r-l)!(m-n+k-r+l)!}.
\end{eqnarray}
    Figure (2) shows that squeezing occurs in both $p$ or
    $x$-quadrature depending on the values of $\xi$ and
    $\lambda$. For $\lambda \leq 1$ only the $p$-quadrature is
    squeezed over all ranges of $|\xi|$ (except near $\xi = 1$)
    and the strength of squeezing is not so much sensible to the
    value of $\lambda$ but depends on the values of $|\xi|$. By increasing $\lambda$-parameter, the
    squeezing transfer from $p$ to $x$-quadrature sooner and
    sooner with respect to the horizontal axis ($|\xi|$). Figures
    $(3a)$ and $(3b)$ show Mandel parameter, Eq. (\ref {mandel}), as a
    function of $|\xi|$ for different values of $\lambda$ in two basis $|n\rangle$ and $|n\rangle_\lambda$.
    According to these results, while the $\lambda$-SSs, Eq. (\ref {represSSs}),
    have sub-Poissonian (nonclassical) behavior in $|n \rangle_
    \lambda$ basis (figure $3a$), it has supper-Poissonian
    statistics in  $|n \rangle _{_{SHO}}$ basis (figure $3b$).
\section{{\bf Conclusion }}\label{sec-ss}
   The states proposed by J-Beckers {\it et al} [Beckers(1998)] (or by H.
   Fu {\it et al}[Fu(2000)]) $|n \rangle_ \lambda$ are not coherent
   ones, but they exhibit squeezing. We showed that these states can
   be regarded as $f$-deformed coherent bound states and demonstrated that our infinite dimensional Hilbert
   space may be spanned by them. Our motivation for this
   consideration is the more generality and more flexibility  of the
   nonorthogonal basis $\{|n\rangle_\lambda, n\in\mathbf{N}\}_{n=0}^\infty$ rather than orthogonal one
   $\{|n\rangle, n\in\mathbf{N}\}_{n=0}^\infty$.
   Then we concluded that by
   some special superposition of the deformed Fock space
   ($|n \rangle_ \lambda$), we can obtain the representations of
   coherent state $|\alpha, \lambda \rangle$ as well as squeezed coherent state
   $| \xi, \lambda \rangle$ in the new bases. In comparison of our
   new nonlinear-nonorthogonal but normalized Fock space with basis
   $|n\rangle_\lambda$ and the orthonormal basis $|n
   \rangle$ of the old Fock space, it is worthable to note that the
   one we consider is not just a change of basis. By this we mean
   that there is a cut-off in the sum on the rhs of relation (\ref {exp-nlamb-n}).
   In other words the sum does not goes on to infinity. So it is not
   surprising that in the new Fock space we gain a set of new
   physical aspects.
   It will be interesting to study other general systems and see
   if their Fock spaces can be constructed by other superpositions
   of number states and then derive new coherent and
   squeezed states.
\section{{\bf Appendix}\\
      How different deformations affect coherent states}\label{appen}
      Here we bring some examples to show that how the deformation
      of creation, annihilation or both of them for a specific dynamical system affect the produced CSs. We
      construct the Hamiltonian
\begin{eqnarray}\label{}\nonumber
  H'=&H_{_{SHO}}^2=(a^\dag a +\frac {1}{2})^2\\
    =&a^\dag (a^\dag a +2)a +\frac {1}{4}\\  \nonumber
    =&a^\dag (\hat{n} +2)a +\frac {1}{4}.
\end{eqnarray}
     Before discussing the coherent states of this nonlinear
     Hamiltonian, let us give some physical aspects for it. First
     it is obvious that $H'|n\rangle= (n+\frac{1}{2})^2 |n\rangle$ which
     means that the states are not equally spaced. We know that
     a number of quantum mechanical systems e.g. square-well [Styer(2001)],
     Morse potential [Morse(1926)] and P\"oschl-Teller potential [Posh(1933)]
     exhibit quadratic spectra $ E_n \sim n^2$, which indicate the
     nonlinearity nature of them. Secondly we bring briefly the physical
     meaning of the above Hamiltonian as Man'ko {\it et al} proposed
     [Manko(1997)]. For this purpose we may apply the inverse procedure they
     proposed, {\it i.e.} replacing creation and annihilation operators
     by complex variables $\alpha$ and $\alpha^\ast$, after all we
     have
\begin{equation}\label{ham}
  H'(\alpha, \alpha^*)=(2+\alpha\alpha^*)\alpha\alpha^*
\end{equation}
     where $\alpha=(x+iy)/\sqrt 2$ and $\alpha^ \ast =(x-iy)/\sqrt
     2, \{x, y\}=1$ and by $\{. , .\}$ we mean the Poissonian
     bracket. Therefore we obtain
\begin{equation}\label{alfa}
   \dot{\alpha}+ i (2+4\alpha \alpha ^\ast)\alpha=0.
\end{equation}
    Since $\alpha \alpha^ \ast$ is a constant of motion we conclude
    that the frequency of this nonlinear Hamiltonian dynamics is
\begin{equation}\label{}
    \omega= 2(1 + 2\alpha \alpha^*).
\end{equation}
     This shows the number of particles dependence of new
     frequency, which is a nonlinear phenomena have been called by
     V. I. Man'ko and G. Tino {\it "the blue shift of the frequency of
     light"}[Manko(1995)].

     Now we try to construct CSs with three different
     deformations imposes to this Hamiltonian as follows:

{\bf I)} $A=f_1(\hat{n}) a, A^\dag=a^\dag $ where
     $f_1(\hat{n})=\hat{n}+2$. By the known procedure we have:
\begin{equation}\label{pp}
      |\alpha, f_1\rangle = C_0 \sum_{n=0}^{\infty} \frac {\alpha
      ^n}{(n!)^{\frac{3}{2}}} |n\rangle,
      \qquad
      C_0 = \left[ \sum _{n=0}^\infty \frac {|\alpha|^{2n}}
      {(n!)^3}\right]^{-\frac{1}{2}}.
\end{equation}
{\bf II)} $B=a f_2(\hat{n}), B^\dag = a^\dag f_2(\hat{n})a$ where
      $f_2(\hat{n})=
      \sqrt{\hat{n}+2}$. In this case we obtain:
\begin{equation}\label{qq}
      |\alpha, f_2\rangle = C_0 \sum_{n=0}^{\infty} \frac {\alpha
      ^n}{n!} |n\rangle,
      \qquad
      C_0 = \left[ \sum _{n=0}^\infty \frac {|\alpha|^{2n}}
      {(n!)^2}\right]^{-\frac{1}{2}}.
\end{equation}
{\bf III)} $C=a, C^\dag=a^\dag f_3(\hat{n}),
      f_3(\hat{n})=\hat{n}+2$ where it is really expected that by
      $C$ operator we demand the usual canonical coherent states.
      Therefore in conclusion it is not surprising that when we
      translate the $\lambda$-CSs, Eq. (\ref{lambda-CS}) in the standard Fock
      space we obtain the canonical CS, clearly because we did not
      change the form of the annihilation operator. As a result we observe that
      different deformations lead to different CSs. The first two
      of these CSs (and obviously the third) are special kinds of
      generalized CSs proposed by K. A. Penson and A. I. Solomon
      [Penson(2001)] in the form:
\begin{equation}\label{}
    |Z\rangle_{_C}=\N_{_C}(|Z|^2)^{-1/2}\sum_{n=0}^{\infty}\frac{Z^n}{\sqrt{C(n)}}
    |n\rangle,\qquad Z\in \mathcal{C},
\end{equation}
     where the normalization factor is
\begin{equation}
   \N_{_C}(|Z|^2)
   = \sum _{n=0}^{\infty} \frac {|Z|^{2n}} {C(n)}.
\end{equation}
\ack {The authors would like to thank Dr. M. H. Naderi for useful
comments and discussions, also for reminding some references.}

 \vspace{1cm}
 {\bf FIGURE CAPTIONS:}

       {\bf FIG. 1.} Mandel parameter for $\lambda-$CSs in
       $|n\rangle_\lambda$ basis as a function of $\lambda$ for
       different values of $\alpha$. $\alpha$ is taken to be real.

       {\bf FIG. 2.} Uncertainty in field quadrature $p$, $(\delta
       p)^2$, as a function of $|\xi|$ for different values of
       $\lambda$.

       {\bf FIG. 3a.} Mandel parameter of $\lambda$-SSs as a
       function of $|\xi|$ for different values of
       $\lambda$ in $|n\rangle_\lambda$ basis.

       {\bf FIG. 3a.} Mandel parameter of $\lambda$-SSs as a
       function of $|\xi|$ for different values of
       $\lambda$ in $|n\rangle$ basis.


\section*{References}
\begin{harvard}
     \bibitem[Ali(2004)]{Ali} S. T. Ali, R. Roknizadeh and M. K. Tavassoly, accepted
              for publication in {\it J. Phys. A: Math. Gen.}{\bf
              37} 4407

     \bibitem[Ali(1993)]{Ali} S. T. Ali, J-P. Antoine and J-P. Gazeau, {\it Ann.
              Phys.} {\bf 222} 1(1993), {\it ibid} {\it Ann. Phys.}
              {\bf 222} 38(1993).

      \bibitem[Bardek(2000)]{Bardek} V. Bardek and S. Meljanace, {\it Eur. Phys. J.}
             {\bf C 17} 539(2000).

      \bibitem[Beckers(1998)]{Becker} J. Beckers, N. Debergh, F. H. Szafraniec, {\it Phys.
              Lett.} {\bf A 243} 256(1998). J. Beckers, J. F. Carinena, N. Debergh and G.
              Marmo, {\it Mod. Phys. Lett.}  {\bf A 16} 91(2001), X-G Wang and H. Fu,
              {\it Mod. Phys. Lett.} {\bf B 14} 243(2000).

     \bibitem[Das(2002)]{Das}P. K. Das, {\it Int. J. Theo. Phys.} {\bf 41},
             1099(2002).

     \bibitem[Daubechies(1986)]{Daubechies} I. Daubechies, A. Grossman and Y. Meyer, {\it J. Math. Phys.}
             {\bf 27} 1271(1986).

     \bibitem[Daubechies(1990)]{Daubechies} I. Daubechies {\it IEEE Trans. Inform. Theory}, {\bf 36}
              961(1990).

     \bibitem[Duffin(1952)]{Duffin} R. J. Duffin and A. C. Schaeffer, {\it Trans. Amer. Math.
             Soc.} {\bf 72} 341(1952).

     \bibitem[Fu(2000)]{Fu} H. Fu, Y. Feng and A. I. Solomon, {\it J. Phys. A: Math. Gen.} {\bf 33},
              2231(2000).

     \bibitem[Klauder((1985)]{Klauder} J. R. Klauder and B. S.
              Skagerstam, {\it "Coherent States"}, World Scientific,
              Singapore(1985).

     \bibitem[Lynn(1986)]{Lynn} P. A. Lynn, {\it "Introduction to the Analysis and Proceeding of Signals"},
             $3^{th}$ ed., Macmillan, London(1986).

     \bibitem[Manko(1995)]{Manko} V. I. Man'ko, G. M. Tino, {\it Phys. lett.} {\bf A 24} 202(1995).

     \bibitem[Manko(1996)]{Manko} V. I. Man'ko, G. Marmo, E. C. Sudarshan and
              F. Zaccaria, {\it "f-oscillators and nonlinear coherent states",
              Proc. $4^{th}$ Wigner Symp.} ed N. M. Atakishiyev, T. H.
              Seligman and K. B. Wolf (Singapore: World
              Scientific) 421(1996).

     \bibitem[Morse(1926)]{Morse} P. M. Morse, {\it Phys. Rev.} {\bf 34} 57(1926).

     \bibitem[Penson(2001)]{Penson} K. A. Penson and  A. I. Solomon, {\it "coherent states
              from combinational sequences"}, $2^{th}$ Conference on Quantum Theory and Symmetry, Cracow, Poland
              17(2001).

     \bibitem[Perelomov(1986)]{Perelomov} A. Perelomov, {"Generalized Coherent States and Their
              Applications"}, Springer-Verlag, New York(1985).


      \bibitem[Peres(1995)]{Peres} A. Peres, {\it "Quantum Theory: Concepts and
              Methods"}, published by Kluwer Academic Publisher(1995).


     \bibitem[Posh(1933)]{Posh} G. Poschl and E. Teller, {\it Z. Phys.} {\bf 83} 143(1933).

     \bibitem[Re'camier(2003)]{Re'camier} J. Re'camier and R. Jauregui, J. Opt. B: Quantum
              Semiclass. Opt. {\bf 5} S365(2003).


     \bibitem[Shanta(1994)]{Shanta} P. Shanta, S. Chaturvedi, V. Srinivasan and R. Jagannathan, {\it
             J. Phys. A: Math. Gen.} {\bf 27} 6433(1994), {\it and}  P. Shanta, S. Chaturvedi, V. Srinivasan,
             G. S. Agarwal and C. L. Mehta, {\it
             Phys. Rev. Lett.} {\bf 72} 1447(1994).

     \bibitem[Sivakumar(2002)]{Sivakumar} S. Sivakumar, {\it J. Opt. B: Quantum Semiclass. Opt.} {\bf
             2}  R61(2002).

     \bibitem[Solomon(1990)]{Solomon} A. I. Solomon and J. Katriel, {\it J. Phys. A: Math. Gen.} {\bf
             23} 5L1209(1990).

     \bibitem[Styer(2001)]{Styer} D. F. Styer, {\it Am. J. Phys.} {\bf 69} 56(2001).

     \bibitem[Wang(2000)]{Wang} X. Wang, {\it J. Opt. B: Quantum Semiclass. Opt.} {\bf 2}
              534(2000).
\end{harvard}
    
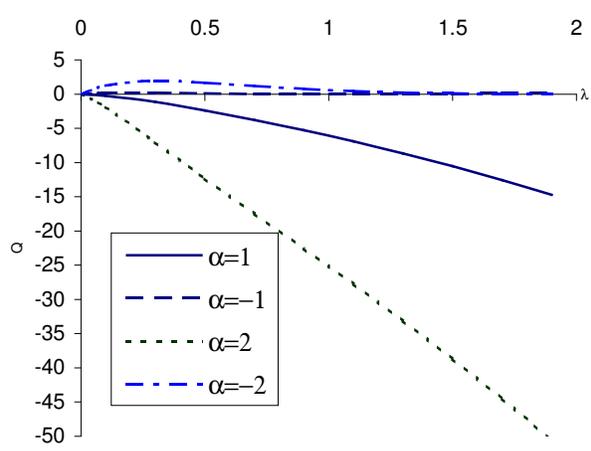
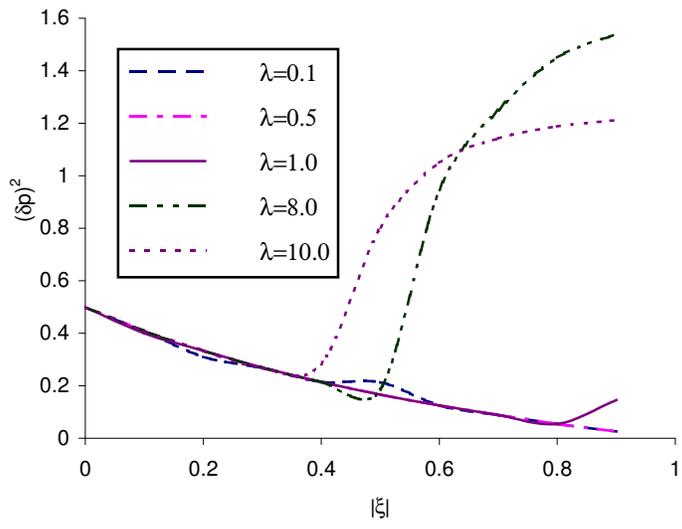
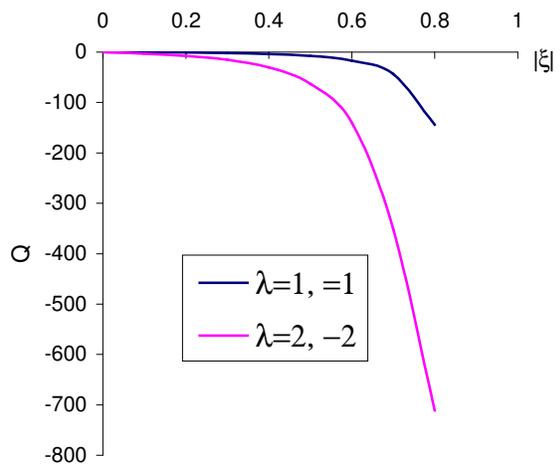
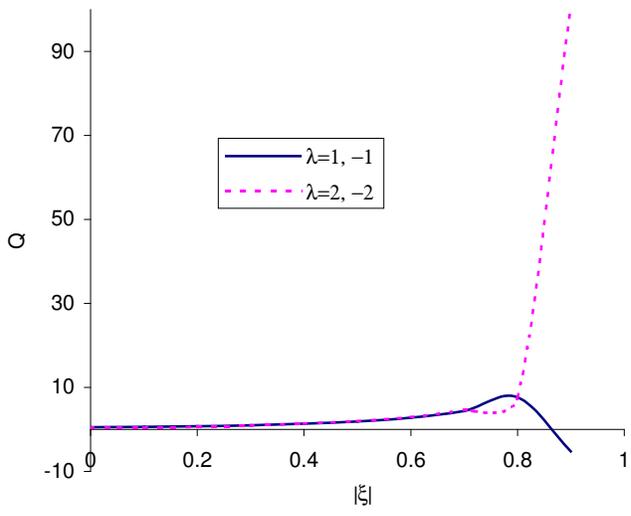
\end{document}